# The Effect of a Strong Stellar Flare on the Atmospheric Chemistry of an Earth-like Planet Orbiting an M dwarf

(Astrobiology, accepted)

Antígona Segura[1,*], Lucianne Walkowicz[2,*], Victoria Meadows[3,*], James Kasting[4,*], Suzanne Hawley[5]

[1]Instituto de Ciencias Nucleares, Universidad Nacional Autónoma de México, [2]University of California at Berkeley, [3]University of Washington, [4]Pennsylvania State University, [5]University of Washington

*Members of the Virtual Planet Laboratory Lead Team of the NASA Astrobiology Institute.

To whom correspondence should be directed:
Antígona Segura
Universidad Nacional Autónoma de México
Instituto de Ciencias Nucleares
Circuito Exterior C.U. A.Postal 70-543 04510 México D.F.
Phone: 52 (55) 5622 4739 ext. 269
Fax 52 (55) 56 22 46 82
E-mail: antigona@nucleares.unam.mx

Running title: Flare effect on an Earth-like planet

**Abstract**

Main sequence M stars pose an interesting problem for astrobiology: their abundance in our galaxy makes them likely targets in the hunt for habitable planets, but their strong chromospheric activity produces high energy radiation and charged particles that may be detrimental to life. We studied the impact of the 1985 April 12 flare from the M dwarf, AD Leonis (AD Leo), simulating the effects from both UV radiation and protons on the atmospheric chemistry of a hypothetical, Earth-like planet located within its habitable zone. Based on observations of solar proton events and the *Neupert effect* we estimated a proton flux associated with the flare of $5.9 \times 10^8$ protons cm$^{-2}$ sr$^{-1}$ s$^{-1}$ for particles with energies >10MeV. Then we calculated the abundance of nitrogen oxides produced by the flare by scaling the production of these compounds during a large solar proton event called the "Carrington event". The simulations were performed using a 1-D photochemical model coupled to a 1-D radiative/convective model. Our results indicate that the ultraviolet radiation emitted during the flare does not produce a significant change in the ozone column depth of the planet. When the action of protons is included, the ozone depletion reached a maximum of 94% two years after the flare for a planet with no magnetic field. At the peak of the flare, the calculated UV fluxes that reach the surface, in the wavelength ranges that are damaging for life, exceed those received on Earth during less than 100 s. Flares may therefore not present a direct hazard for life on the surface of an orbiting habitable planet. Given that AD Leo is one of the most magnetically-active M dwarfs known, this conclusion should apply to planets around other M dwarfs with lower levels of chromospheric activity.

**Introduction**
Although initially deemed unlikely to support habitable planets (Dole, 1964), main sequence M stars (dM, MV, M dwarfs or red dwarfs) are currently enjoying a renaissance as potential targets in the search for habitable planets and life beyond our Solar System (Scalo et al., 2007; Tarter et al., 2007). M dwarfs are the most abundant and long-lived stars in the Galaxy, comprising 70% of the stars in the Solar Neighborhood (Bochanski 2008). Their main sequence lifetimes are much longer than $10^{10}$ years (more than the present calculated age of the Universe). The prevalence of M dwarfs in the Galaxy, combined with their small stellar masses and radii, increases the likelihood of detecting orbiting, terrestrial planets around them through radial velocity, transit, or gravitational microlensing techniques. The M3 dwarf Gl 581 hosts the best known candidates for habitable, terrestrial planets: the recently detected Gl 581 c and d (von Bloh et al., 2007; Selsis et al., 2007).

However, M dwarfs pose unique challenges to habitability of their attendant planets[1]. Because M dwarfs are low luminosity stars, their habitable zones (as defined by Kasting et al. 1993) lie at small orbital radii (~0.2 AU or less), potentially resulting in strong tidal effects on a planet within that zone. It was once thought that the atmospheres of tidally locked planets around M dwarfs would condense on the permanent night side and subsequently collapse. However, Haberle et al. (1996) showed with a 1-D energy balance model that only ~0.1 bar of $CO_2$ is required to maintain sufficient heat transport to ensure atmospheric stability. This result was confirmed by Joshi et al. (1997) using a 3-D general circulation model, and again recently by Joshi (2003) and Edson et al. (2010) with more sophisticated 3-D models. Tidal effects may also influence planetary habitability by causing the orbits of planets close to their host star to evolve, such that planets may move either into or out of the habitable zone over time. Barnes et al. (2007) showed that planets formed with eccentricities great than 0.5 may have shortened habitable lifetimes as a consequence of their orbital evolution. However, they also conclude that because the orbits of terrestrial planets around low mass stars circularize quickly (in < 1Gyr), these planets may not suffer from shortened habitable lifetimes.

Planets in the habitable zones of M dwarfs should also be affected by stellar activity. Although M dwarfs emit the bulk of their flux in the optical and near infrared, many of these stars exhibit magnetic activity that produces high energy charged particles and radiation at short wavelengths, from X-rays to ultraviolet (UV), that may be dangerous for life. Strong stellar activity also affects the retention of planetary atmospheres by driving thermal and non-thermal atmospheric escape processes. A critical question for habitability is whether planets around M dwarfs are able to retain their atmospheres, given the strong activity of their parent stars. UV emission heats the outermost layer of the planetary atmosphere (the "exosphere") and dissociates molecular species, contributing to mass loss. If activity-driven escape processes are efficient, H/He envelopes may be stripped from planets around M dwarfs. This loss may be detrimental to planets with thin atmospheres, but it may also erode gas giants into volatile-rich, Neptune-mass cores (Baraffe et al. 2005; Lammer et al. 2007). These published atmospheric loss calculations for M-star planets have all been done using hydrostatic model atmospheres. Paradoxically, this may *overestimate* atmospheric loss by allowing these atmospheres to have extended cross sections that interact strongly with stellar winds. Hydrodynamic upper atmosphere models, *e.g.*, Tian et al. (2008), are needed to check these simulations.

Stellar activity is of particular concern for the *continuity* of habitability on the planetary surface, where starspots or flares may cause the stellar irradiance to vary with time. In an early exploration of this issue, Heath et al. (1999) concluded that as the UV flux from even strongly active M dwarfs is typically much less than the solar UV irradiance on Earth, such emission is unlikely to pose a threat to habitability of planets around these stars. More recently, Buccino et al. (2006) considered the effects of stellar UV variability on the location of the habitable zone and the formation of complex biogenic molecules on the planet. However, there has not yet been a detailed, dynamic model exploring the evolution of an Earth-like atmosphere over the course of a flare, and so whether stellar flares present a "show-stopping" barrier to planetary surface habitability remains an outstanding question.

---

[1] For a complete review of the potential habitability of planets around red dwarfs we refer the readers to the reviews by Scalo et al. (2007) and Tarter et al. (2007).

During a stellar flare, magnetic reconnection events in the outer atmosphere of the star accelerate energetic electrons downward along magnetic field lines, where they impact and heat the lower atmosphere and produce hard X-rays, increased chromospheric line emission, and enhanced continuum emission. The initial injection of energy from the flare into the stellar atmosphere creates a rapid initial rise and peak in photometric brightness, known as the impulsive phase, followed by a long tail in energy output while the star slowly returns to its quiescent state, known as the gradual phase. The impulsive phase of the flare is mainly characterized by the continuum increase, followed shortly thereafter by enhancement in the stellar emission lines. In M dwarf flares, the continuum enhancement appears as a ~9000-10,000K blackbody superimposed on the spectrum of the star, and comprises the majority of the energy release during the early stage of the flare. As the flare progresses, this initial blue continuum increase fades, but the enhanced emission lines persist, dominating the energy output during the gradual phase of the flare. The impulsive phase is generally much shorter than the gradual phase, lasting on the order of minutes to half an hour, while the gradual phase is typically several hours in duration for a large flare. Because of the strong blue continuum increase associated with large flares, the slope of the stellar spectrum changes dramatically over the course of the flare.

Correctly treating the energy input to a planetary atmosphere from a stellar flare requires broad wavelength coverage observations during the flare to account for the time and spectral dependence from the UV (1000 Å) to the visible (5000 Å). Most flares have similar light curve shapes (*e.g.* Welsh et al. 2007) but their wavelength variation with time is known only for some parts of the spectrum, usually either X-rays or the visible, as broad wavelength coverage observing campaigns tend to be rare. There are just a few observations of flares that simultaneously cover the full 1000 to 5000 Å spectral range (Elgarøy et al. 1988, Hawley & Pettersen 1991, van den Oord et al. 1996, Jevremović et al. 1998, Montes et al. 1999, Hawley et al. 2003); therefore generalizations are hard to make. The goal of this work is to understand how the ozone concentration profile of the planetary atmosphere changes during a flare, given that ozone is one of the best compounds for planetary biosphere detection via remote sensing (e.g. Des Marais et al. 2002) and that it can protect life on the planetary surface from potentially damaging UV radiation. As empirical input, we use the observations by Hawley and Pettersen (1991) of the great AD Leonis flare of 1985. AD Leonis (AD Leo, Gl 388) is a dM3e star located at 4.85 pc from the Sun. Its ratio of soft X-ray to bolometric luminosity, which can be considered as a measure of its activity, is $10^3$ times larger than that of the Sun (Fig. 5, Scalo et al. 2006). Because of its proximity, AD Leo is a relatively bright star, with a visual magnitude V=9.43. As a consequence of its brightness and its high flare rate, it is one of the most observed flare stars. Based on solar observations we estimated the fluence of a proton event associated to the AD Leo flare. Protons that arrive at Earth's atmosphere during a solar flare produce an enhancement of odd nitrogen (N, NO, $NO_2$) and odd hydrogen (H, OH, $HO_2$) in the upper stratosphere and mesosphere. Both chemical species destroy ozone through catalytic reactions. In this paper, we use a convective/radiative climate model coupled to a photochemical model to simulate the time-dependent atmospheric effects of an energetic stellar flare on an Earth-like planet located within the habitable zone of AD Leo. Because flares can occur without proton events we explore both the atmospheric effects of UV radiation alone, and the combined effects of UV radiation and protons from the flare on the atmospheric composition and surface radiation environment.

**Planetary atmosphere simulation**
We simulated the effect of a stellar flare on the atmosphere of an Earth-like planet located within AD Leo's habitable zone at 0.16 AU. That is the 1-AU equivalent distance for AD Leo, the distance where the planet receives the same integrated energy from AD Leo as Earth does from the Sun (Segura et al, 2005). Because the simulated atmosphere contains high concentrations of methane, we adjusted the semi-major axis from 0.1595 to 0.1603 AU to give a planetary surface temperature of ~288K. To model the evolution of the planet's atmosphere during the flare, we modified two existing atmospheric modeling codes to work in a coupled, time-dependent manner. The codes were a radiative-convective model (Pavlov et al. 2000) and a photochemical model (Segura et al. 2003; Segura et al. 2005). Improved versions of these models were coupled in a time-dependent mode to calculate the changes in temperature, water content and chemical composition of an Earth-like planetary atmosphere composed of 0.21 $O_2$ and 0.78 $N_2$. The codes were calibrated so as to closely reproduce the 1976 U.S. Standard Atmosphere, given the conditions of present Earth irradiated by the present Sun (Fig . 1).

The radiative–convective model is actually a hybrid of two separate models. The time-stepping procedure and the solar (visible/near-IR) portion of the radiation code are from the model of Pavlov et al. (2000). The solar code incorporates a δ two-stream scattering algorithm (Toon et al., 1989) to calculate fluxes and uses four-term, correlated-$k$ coefficients to parameterize absorption by $O_3$, $CO_2$, $H_2O$, $O_2$, and $CH_4$ in each of 38 spectral intervals (Kasting and Ackerman, 1986). At thermal-IR wavelengths, we used the independent rapid radiative transfer model (RRTM) implemented by Segura et al. (2003), based on an algorithm developed by Mlawer et al. (1997). RRTM uses 16-term sums in each of its spectral bands in which the $k$-coefficients are concentrated in the areas of most rapidly changing absorption, thereby providing better spectral resolution at altitudes where Doppler broadening is important. Spectral intervals and included absorber species are described in Mlawer et al. (1997), which fully details the method. A disadvantage of this version of the model is that the $k$-coefficients used therein are not applicable to dense, $CO_2$-rich atmospheres. This limitation is not a problem for the current study because we have restricted our calculations to relatively $CO_2$-poor atmospheres like that of the modern Earth. The most recent version of RRTM (http://rtweb.aer.com/) has been validated up to concentrations of 100 times current $CO_2$. The (log pressure) grid extended from the assumed surface pressure of 1 bar down to $10^{-5}$ bar. The program subdivided this range into 52 levels. Interpolation was required between the climate code and the photochemical code, which ran on a fixed altitude grid. The code worked with a variable time-stepping procedure. Starting from an established value, the code calculated the ensuing time steps based on the difference between the last two temperature profiles: the larger the difference, the smaller the time step. Before the flare the initial time step is $10^4$ s, during the flare it is 50 s, after the flare it is 500 s.

The photochemical model, originally developed by Kasting et al. (1985), is detailed in Segura et al. (2003). It solves for 55 different chemical species that are linked by 217 separate reactions. The altitude range extended from 0 to 64 km in 0.5-km increments. Photolysis rates for various gas-phase species were calculated using a δ two-stream routine (Toon et al., 1989) that accounts for multiple scattering by atmospheric gases and by sulfate aerosols. The model uses the (fully implicit) reverse Euler method to time step to a solution. The initial time step is $10^{-4}$ s, and this increases automatically as different species reach equilibrium. All simulations were performed with a constant planetary surface pressure of 1 atm. The $CO_2$ mixing ratio was kept constant at 355 ppmv. Argon was maintained at 1% of the total atmospheric composition. The photochemical model was run using a fixed solar zenith angle of 42°, as this was found to best reproduce Earth's ozone column depth of 0.32 atm-cm ($8.61 \times 10^{18}$ $cm^{-2}$) reported by McClatchey et al. (1971). Our calibration of the model resulted in an ozone column depth for Earth around the Sun of $8.25 \times 10^{18}$ $cm^{-2}$ (0.307 atm-cm). Calculated photolysis rates were multiplied by 0.5 to account for the diurnal cycle. This is a crude approximation for tidally locked M-star planets; however, we assume that atmospheric mixing between the day- and night-sides would result in some sort of similar averaging. The radiative–convective model used the daytime average solar zenith angle of 60° and the same factor of 0.5 for diurnal variation. The planetary surface albedo was set at 0.2, a number that allows the model to reproduce the temperature profile of present Earth even though clouds are omitted (for a more detailed discussion, see Segura et al. 2003). The boundary conditions for the photochemical model were fixed surface fluxes for $CH_4$, $N_2O$ and $CH_3Cl$, and fixed deposition velocities for $H_2$ and CO. For $CH_4$, a fixed, relatively low surface flux was used to avoid "methane runaway", an unrealistic accumulation of methane in the atmosphere. This assumption is to some extent arbitrary; however, it is expected to have little effect on the simulations (see below).

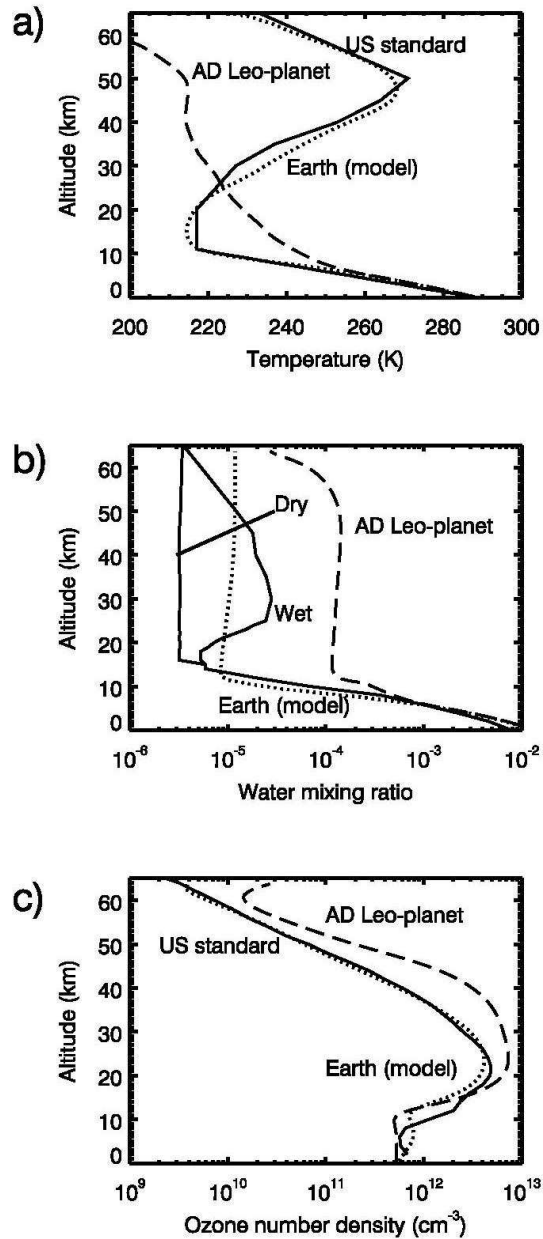

Figure 1. Model results for present Earth around the Sun (dotted lines), compared with vertical profiles from the 1976 U.S. Standard Atmosphere (solid lines) and the profiles for the AD Leo planet (dashed lines): (a) temperature, (b) H2O mixing ratios, and (c) O3 number density. The profiles of the AD Leo planet are the ones calculated for steady state and used as staring point in the simulations.

On the present Earth, the mechanism that destroys methane in the troposphere is:

$$O_3 + h\nu \ (\lambda < 3100 \text{ Å}) \rightarrow O_2 + O(^1D)$$
$$O(^1D) + H_2O \rightarrow 2 \text{ OH}$$
$$CH_4 + OH \rightarrow CH_3 + H_2O$$
$$CH_3 + O_2 + M \rightarrow CH_3O_2 + M \rightarrow \ldots$$
$$\rightarrow CO \text{ (or } CO_2) + H_2O$$

This process depends heavily on the amount of ozone available in the atmosphere, and how rapidly the ozone is photolyzed to from $O(^1D)$. Ozone is produced primarily via the photolysis of oxygen, which occurs at wavelengths <2000 Å. Ozone itself is photolyzed at wavelengths between 2000 and 8000 Å. In quiescence, AD Leo's UV flux delivers more energy to its planet at 1000-2000 Å than the Sun does to the Earth (see Fig. 2). Consequently, more $O_2$ is photolyzed and more $O_3$ is produced on the AD Leo planet. The total ozone column depth of the AD Leo planet obtained for this simulation was $1.67 \times 10^{19}$ cm$^{-2}$, while Earth's ozone column depth calculated by the same model is $8.25 \times 10^{18}$ cm$^{-2}$. In the 2000-3500 Å wavelength range, the opposite happens: AD Leo delivers less energy to its planet than the Sun does to Earth (Fig. 2), and so $O_3$ is photolyzed less rapidly. As a result, there are fewer $O(^1D)$ atoms and, accordingly, fewer OH molecules to react with methane. Thus, while our model produces an OH number density of $\sim 10^6$ cm$^{-3}$ for present-day Earth, the corresponding value for the AD Leo planet was $\sim 10^2$ cm$^{-3}$. The methane surface flux to reproduce the present Earth methane concentration (1.6 ppm) is calculated by the model to be $7.3 \times 10^{14}$ g/yr. If we use this number as boundary condition for methane, $CH_4$ accumulates to concentrations >500 ppm—too high to be accurately simulated with our RRTM-based climate model. We choose a constant methane surface flux of $6.4 \times 10^{14}$ g/yr, or about 88 percent of the terrestrial value listed above. Given this methane influx, the photochemical code calculates an atmospheric $CH_4$ concentration of 334 ppmv.

To simulate the effect of a stellar flare on a planetary atmosphere, we started from a steady-state solution obtained as described in Segura et al (2003, 2005) using the AD Leo spectrum during quiescence (black solid line in Fig 2). Once the steady solution was reached, the input stellar flux for wavelengths shorter than 4400 Å was changed to that of the stellar flare at a given time. The flare time steps correspond to changes on the input stellar flux (Fig. 2) and vary between 91 s to 371 s, depending on the time intervals between the available flare spectra. Given a flare time interval, each code ran a different number of time steps according to its own setup and solution method. For example, the first flare interval ran from 0 to 100 seconds, and its initial conditions are those for the steady state. The photochemical model was run with variable time steps that started at $10^{-4}$ s and were increased as the chemical species reached equilibrium. Once the photochemical model reached the flare time step of 100 s, the outputs were saved, and these were passed on to the climate model. The climate model used an initial time step of 50 s, which was then increased until it reached 100 s, using a variable time step procedure. The outputs of both models at the end of a given flare interval were used as initial conditions for the next time interval. Once the flare ended, the subsequent evolution of the atmosphere was tracked until it reached steady state. We followed an algorithm similar to that used during the flare, but the AD Leo flux was held constant in the quiescent state. This allowed larger time steps to be used during the recovery phase.

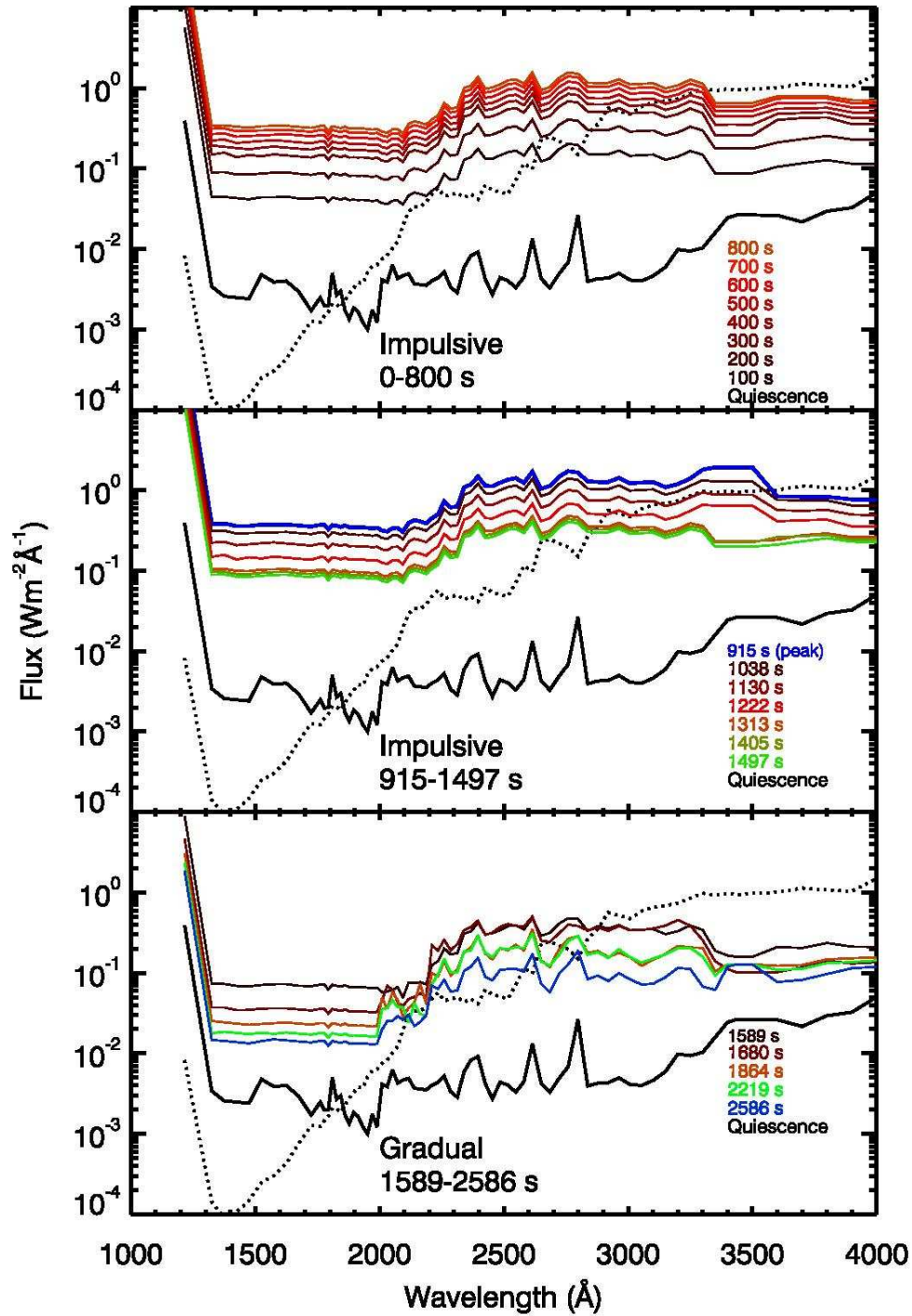

Figure 2. Flux received at the top of the atmosphere of a planet on the AD Leo habitable zone (0.16 AU). The dotted line is the solar flux received by Earth. Times listed on the right lower corner of each panel correspond to the flare fluxes plotted on that panel. AD Leo spectrum during quiescence is always shown in a black continuous line to be used as a reference.

**Model Input Data**
*Input Stellar Flare Spectra*
The time-dependent sequence of stellar UV/visible input to the atmospheric model was constructed from the multiwavelength flare observations of Hawley & Pettersen (1991). This campaign observed a large flare (~$10^{34}$ ergs in total energy, with a duration of over 4 hours) on AD Leo, obtaining photometry and optical and UV spectra over the duration of the flare. Optical spectra were obtained with the McDonald Observatory 2.1 m telescope and covered the wavelength range of 3560-4440 Å, both during quiescence and over the flare duration. UV spectra were obtained contemporaneously with the International Ultraviolet Explorer (IUE) satellite. Two far-UV spectra were obtained with the IUE's short wavelength (SWP) camera; these spectra extend from 1150 to 2000 Å and were obtained during quiescence and during the flare, respectively. Five flare and seven quiescent near-UV spectra were obtained with the long wavelength (LWP) camera, covering 1900 to 3000 Å. A complete description of the observing method and data reduction can be found in Hawley & Pettersen (1991).

Optical spectra were taken over the entire duration of the flare and during quiescence, and were therefore available for every stage of the flare. UV spectra were unfortunately not as well distributed in time, with only one short-wavelength spectrum and five long-wavelength spectra taken during the flare. In addition, many of the UV emission lines were saturated in the IUE UV spectra.

The boundary between the impulsive and gradual phase of the flare is defined to be at the inflection point in the photometric flare light curve, where the slope of the light curve becomes shallower after its initial steep rise, peak, and rapid decrease during the impulsive phase. In the case of the Hawley & Pettersen (1991) flare, the division between the impulsive and gradual phases was determined to be at ~1600s. Optical spectra were taken during the impulsive phase at roughly 100s intervals, at 915s, 1038s, 1130s, 1222 s, 1313s, 1405s and 1497s. Optical gradual phase spectra were taken at 1589s, 1680s, 1864s, 2219s and 2586s, by which time the 2586s spectrum the blue flare continuum had largely disappeared, leaving only the persistent enhancement in the emission lines.

The IUE UV spectra were contemporaneous with the optical spectra, but unevenly distributed over the flare duration. The five near-UV flare spectra were taken at 1604s, 2542s, 6390s, 8469s, and 13225s. The maximum enhancement of the blue optical continuum took place 915s after the flare start, so we took the optical spectrum at this time as representative of the peak of the flare. As the first IUE near-UV observation did not occur until ~1600s after the flare start, it was necessary to scale this first IUE spectrum to join it with the blue optical spectrum. Using the light curves from Figure 1 of Hawley & Pettersen (1991), we estimated that the increase in the continuum around 2800 Å was ~3.6 times larger at 915s than the continuum observed at 1600s. Whereas the Hawley & Pettersen (1991) flare had two successive peaks in the light curve, we assumed a simpler single peak followed by a gradual decline, such that the spectrum at 915s represented the single peak of our flare. Scaling the long-wavelength end of the near-UV spectrum to join the blue end of the optical spectrum also yielded a factor of 3.6 increase. We then interpolated between this scaled near-UV spectrum and the next near-UV observation at 2542s to generate near-UV spectra to join the optical spectra taken within that time interval. The blue optical spectrum taken at 2586s was joined with the nearly simultaneous near UV spectrum taken at 2542s to complete the sequence.

The near-UV spectra also had to be interpolated to characterize the initial rise during the impulsive phase. To accomplish this, we assumed that the change of the stellar spectrum as the blue continuum declined during the gradual phase was similar in progression to the initial continuum enhancement, but occurred over a shorter time interval. We compared the median flux between 3000 and 3100 Å in the quiescent spectrum to the spectrum from the peak of the flare (915s), and found that the flux in this wavelength interval rose from $1.23 \times 10^{-14}$ ergs cm$^{-2}$ s$^{-1}$ Å$^{-1}$ in quiescence to $3.5 \times 10^{-12}$ ergs cm$^{-2}$ s$^{-1}$ Å$^{-1}$. We then compared the flux levels in this same wavelength interval to that of the gradual phase spectra to select the most representative spectrum for each 100s interval during the initial flare rise. Only two observations were available in the far-UV: one taken during the first 900s of the flare, and one in quiescence. We joined the FUV flare spectrum with the NUV and optical composite spectrum for 915s. We then interpolated between the quiescent and flare spectra to match the NUV-optical composite spectra at the other time steps during the flare.

During the flare, many of the strong emission lines in the IUE UV spectra (e.g. C IV, Ly $\alpha$, Fe II, Si II and Mg II) were saturated. Although the stellar input spectra were rebinned prior to being used as input to the photochemical model, and therefore narrow features such as emission lines were not resolved, we include them here for completeness. To estimate the flux in these lines, we examined the work of Byrne & Doyle (1989) and Doyle et al. (1990) and adopted their values measured for YZ CMi, an M dwarf with an activity level similar to AD Leo. Artificial emission lines with these integrated fluxes were then used to replace the saturated lines in the flare spectra. The resulting series of composite broad-wavelength spectra are shown in Figure 3. Table 1 summarizes the spectra used to create the input spectra at each time step during the flare.

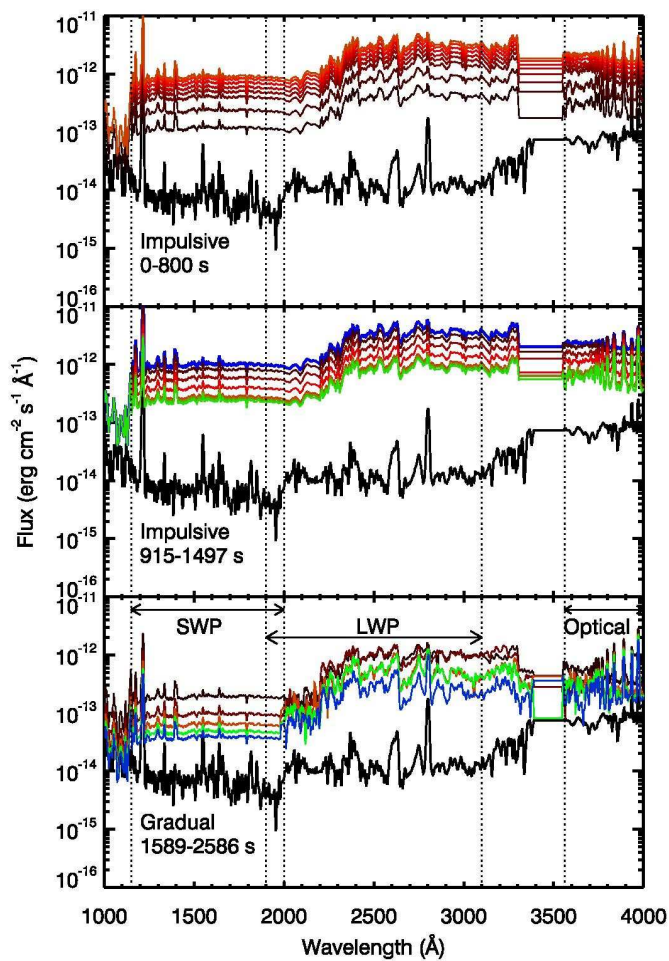

Figure 3. AD Leo flux in quiescent state (black line) and during the flare as observed from Earth. Vertical dotted lines show the wavelength extent of the various spectrograph gratings: the IUE short wavelength SWP grating, from 1150 - 2000Å, the IUE LWP long wavelength grating from 1850 - 3300Å, and the blue edge of the optical spectrum at 3560Å, obtained with the McDonald Observatory Electronic Spectrograph No. 2. IUE spectra were taken in low resolution mode, providing a resolution of 6Å, while the optical spectra have a resolution of 3.5Å. As the bulk flux distribution, rather than the detailed line profiles, are the primary concern in this work, the spectra were rebinned prior to being input to the photochemical/climate model.

Table 1. Flux flare data

| Time (sec) | Optical Spectrum | Near-UV Spectrum | Far-UV Spectrum | $f_{\lambda 2800}$ $\times 10^{-12}$ ergs s$^{-1}$ cm$^{-2}$ Å$^{-1}$ | $f_{\lambda 2000}$ $\times 10^{-12}$ ergs s$^{-1}$ cm$^{-2}$ Å$^{-1}$ |
|---|---|---|---|---|---|
| *Quiescent* | | | | | |
| 0 | quiescent | quiescent IUE | | | |
| *Impulsive Phase* | | | | | |
| 100 | interpolation between quiescent and 1497s | 1604s scaled to match blue end of optical spectrum | interpolation between flare and quiescent, scaled to match near-UV | 0.59 | 0.11 |
| 200 | 1497s | … | … | 1.18 | 0.22 |
| 300 | 1222s | … | … | 1.99 | 0.36 |
| 400 | interpolation between 1222s and 1130s | … | … | 2.46 | 0.45 |
| 500 | 1130s | … | … | 2.92 | 0.53 |
| 600 | interpolation between 1130s and 1038s | … | … | 3.50 | 0.64 |
| 700 | 1038s | … | … | 4.06 | 0.74 |
| 800 | interpolation between 915s and 1038s | … | … | 4.54 | 0.83 |
| 915 | … | … | flare IUE spectrum | 5.02 | 0.92 |
| 1038 | … | … | … | 4.06 | 0.74 |
| 1130 | … | … | … | 2.93 | 0.53 |
| 1222 | … | … | … | 1.99 | 0.36 |
| 1313 | … | … | … | 1.39 | 0.25 |
| 1405 | … | … | … | 1.28 | 0.23 |
| 1497 | … | … | … | 1.18 | 0.22 |
| *Gradual Phase* | | | | | |
| 1589 | … | 1604s IUE spectrum | … | 1.41 | 0.19 |
| 1680 | … | interpolation between 1604s and 2542s | … | 1.18 | 0.66 |
| 1864 | … | … | … | 1.20 | 0.17 |
| 2219 | … | … | | 1.24 | 0.11 |
| 2586 | … | 2542 | … | 0.97 | 0.07 |

*Input Proton Flux and Nitrogen Oxides production*

Simulations of the effect of solar proton events (SPE) on Earth show that Earth's $O_3$ column was likely locally depleted by ~10 percent (Rodger et al. 2008, Thomas et al. 2007) during the 1859 Carrington event, a particularly energetic SPE. This SPE had a calculated fluence of ~$10^{10}$ cm$^{-2}$ for protons with energies >30 MeV, which is roughly four times larger than the solar proton fluence of the largest event from the "spacecraft era" (Rodger et al. 2008).

Given that we are interested in studying the effect of flares on atmospheric biosignatures and the surface radiation environment, we include the effect of proton events that may by associated with flares. Analyses of the observations of solar flares have shown that the probability of detecting proton events increases with the X-ray flare intensity (Belov et al. 2005). This supports the hypothesis that acceleration of protons takes place in the same active regions and at the same time as X-ray flares are produced (Belov et al. 2005). On M dwarfs we are not able to measure proton events, but we can use X-rays as a proxy, at least for the most intense flares. This is supported by observations that show that solar flares and active M dwarfs obey common relationships (Butler et al. 1988). Belov et al. (2005) found that solar X-ray flare intensity can be quantitatively related to proton fluxes:

$$I_p(>10\text{MeV}) = (4.8 \pm 1.3) \times 10^7 \, I_x^{1.14 \pm 0.14} \tag{1}$$

Here, $I_p(>10\text{ MeV})$ is the proton flux for protons with energies larger than 10 MeV in *proton flux units* (pfu = protons cm$^{-2}$ sr$^{-1}$ s$^{-1}$) and $I_x$ is the maximum X-ray intensity in W/m$^2$ of the flare associated with the proton event.

To obtain the proton flux associated with the flare studied here, we calculated the X-ray energy emitted by the big AD Leo flare by using the "Neupert effect" (Neupert 1968; Dennis and Zarro 1993). According to these authors, the radiative loss rate during the gradual phase is directly proportional to the cumulative impulsive energy input, which suggests that the energy input from the non-thermal electrons is responsible for the heating of the plasma. The Neupert effect has been shown to exist between impulsive U-band and soft X-ray emission (Güdel et al. 2002; Hawley et al. 1995) and between near- and mid-UV emitted energy and X-ray luminosity (Mitra-Kraev et al. 2005). Thus, for the big AD Leo flare studied here, we are able to calculate both the X-ray emission and the proton flux associated with the flare.

Mitra-Kraev et al. (2005) found a power-law relation between the flare energy emitted in the UV and the X-ray peak luminosity given by:

$$\mathcal{L}_x = 10^{-4.4} \mathcal{E}_{UV1}^{1.08}$$
$$\mathcal{L}_x = 10^{-15} \mathcal{E}_{UV2}^{1.4} \tag{2}$$

Here, $\mathcal{L}$ is the spectral luminosity density, defined as $\mathcal{L}=L/\Delta$, and $\mathcal{E}$ is the spectral energy density, $\mathcal{E}=E/\Delta$, where $\Delta$ is the width of the respective passband. The subindex UV1 is for the wavelength range 2450–3200 Å, and UV2 stands for the wavelength range 1800–2250 Å.

For the big AD Leo flare studied here we used the values reported by Hawley and Pettersen (Table 6, 1991), converting the integrated fluxes to energies using AD Leo's distance of 4.85 pc. The spectral energy density was calculated using the passband width reported by Hawley and Pettersen (1991). Using these numbers we obtained an X-ray luminosity density of $8.1 \times 10^{28}$ erg s$^{-1}$ Å$^{-1}$ for UV1 and $6.5 \times 10^{27}$ erg s$^{-1}$ Å$^{-1}$ for UV2. For our purpose we are using the largest number in order to obtain the maximum effect on the planetary atmosphere. The spectral X-ray range used to relate solar X-ray flares and protons is 1-8 Å (Livshits et al. 2002); therefore, this number was used to convert the X-ray luminosity density to luminosity. At 0.16 AU the intensity of the X-ray flare in the 1-8 Å wavelength range, is 9 W/m$^2$, so, using Eq. 1, the resultant proton flux is $5.9 \times 10^8$ pfu for protons with energies >10MeV.

To estimate the fluence of the proton event associated with the flare studied here, we assume that the X-ray flare and proton event have the same duration as the UV flare. This is based on observations of flares on M dwarfs (Mitra-Kraev et al. 2005) and solar X-ray flares and their associated proton events (Belov et al. 2005). The fluence for the proton event is $1.5\times10^{12}$ cm$^{-2}$, or 200 times larger than the fluence calculated for the Carrington event. The estimated fluence represents an upper limit considering the various assumptions we have made.

It has been shown that for ionizing photon events of large fluence (which have atmospheric effects similar to protons), the production of nitrogen oxides scales linearly with fluence (Ejzak et al. 2007, Thomas et al. 2007). Based on this, we scale nitrogen oxide production for the Carrington event in this same manner. We use data calculated by Rodger et al (2008), who did simulations using the Sodankylä Ion and Neutral Chemistry (SIC) model (Verronen et al. 2005, Turunen et al. 1996). To include the effect of the proton event in our model we introduce an increase in the number density of $NO_x$ ($\equiv NO + NO_2$) at the peak of the flare (915 s). The increase was proportional to the maximum change calculated at each altitude by Rodger et al. (2008). Changes were introduced at altitudes equal to or larger than 20 km, as the effects of energetic particles on Earth's atmosphere are not important at lower altitudes (Thomas et al. 2007).

In order to test the influence of a possible time lag between the UV flare and the proton event, like the time lag observed between X-ray and UV flares (Mitra-Kraev et al. 2005), we did another simulation in which the increment of $NO_x$ was introduced at 1864 s.

The direct effect of X-rays on the planet was not included here because they are relevant for the upper atmosphere (>70 km) but they do not penetrate to the low stratosphere. They do create NO at high altitudes, ~ 110 km (Barth et al. 1999), but little of this NO is expected to make it down to the stratosphere. Other, secondary effects of this radiation may have some impact on conditions at a planet's surface. Absorption of X-ray energy by molecules creates primary electrons as a photoproduct. These very energetic charged particles then produce secondary photoelectrons which excite other molecules and create aurora-like emission (Smith et al., 2004). These authors calculated the effect of ionizing radiation on a planetary atmosphere using a Monte Carlo code developed to treat Compton scattering and photoabsorption. They showed that X-rays are efficiently blocked by terrestrial atmospheres with interaction depths as small as 0.1 times that of Earth atmosphere. With UV redistribution, a fraction of $2 \times 10^{-3}$ of the incident energy may reach the planetary surface in the 2000–3200 Å region considering $O_2/O_3$ column density profiles similar to present Earth. The X-ray energy estimated for the AD Leo flare studied here is 9 W/m$^2$, so the energy redistributed as UV radiation at the planetary surface should be < 0.018 W/m$^2$. This is an upper limit because the ozone column depth of the AD Leo planet is twice present Earth's ozone column depth. For the AD-Leo planet the UV radiation at the surface in that same wavelength range is 0.014 W/m$^2$ at stellar quiescence and 2.85 W/m$^2$ at the flare peak. For Earth around the Sun the energy that reaches the surface is 3.65 W/m$^2$. This means that X-rays may have a measureable effect on surface UV, but not necessarily a dangerous one, therefore we have not consider X rays on our simulation.

**Results**
*Atmospheric response to UV emitted by flares*
Figures 4 to 7 present the results for the most important atmospheric parameters: temperature, water vapor concentration, ozone number density, and total ozone column depth. During the flare the vertical temperature profile (Figure 4) was perturbed by only a few degrees. At the surface, the maximum change is at the end of the flare, at which point the temperature cools by 0.6 K. This drop was almost overcome during the recovery from the flare as the final temperature difference was only 0.1 K compared to that of the initial steady state. In the stratosphere the maximum temperature change was −5.4 K around 60 km at the end of the flare. During the recovery, the temperature change in the stratosphere (45 to 65 km) varied from 1 to 8 K, reaching the largest difference at higher altitudes. Once the atmosphere reached steady state after the flare, the temperatures above 45 km were about 3 K cooler than those at the same altitudes in the initial steady state.

Water vapor was highly depleted in the upper stratosphere as the flare developed (Fig. 5), reaching its minimum value at the end of the flare with ~0.3 parts per million per volume (ppmv) around 60 km. The initial steady-state values at this altitude were around 30 ppmv. Methane, which has a long photochemical lifetime (~1000 years) in this atmosphere, did not change appreciably during the flare, keeping a concentration of 335 ppmv in the troposphere. The ozone number density (Figure 6) decreased in the stratosphere during the impulsive phase of the flare and then increased until and after the flare ended. The effect of the flare on the total ozone column depth is plotted in Figure 7; the maximum change in ozone number density during the flare is a depletion of 1% seen 1000s after the flare onset.

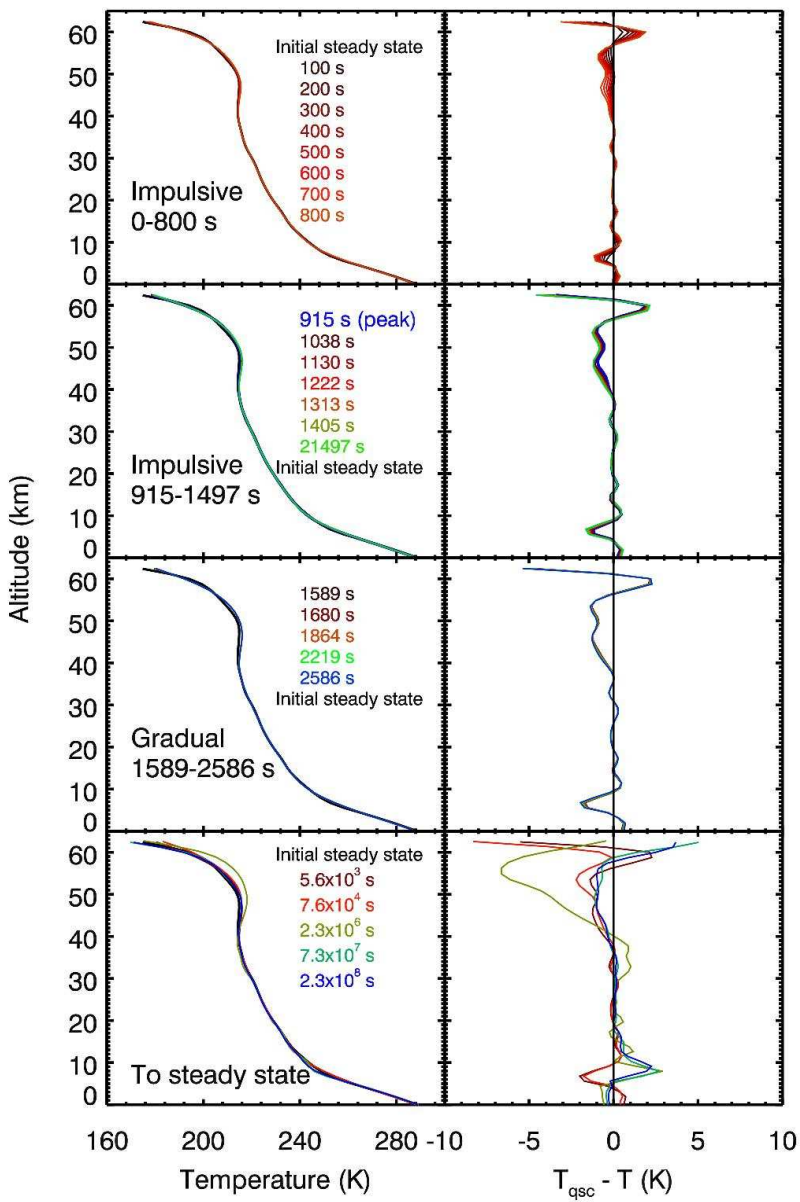

Figure 4. Effect of incident UV radiation from a flare on the temperature profile. Left: Temperature profile for an Earth-like planet around AD Leo, before, during, and after a big UV flare event. Right: Difference between the initial steady state temperature ($T_{qsc}$) and the temperature calculated for the AD Leo planet during and after the big UV flare event.

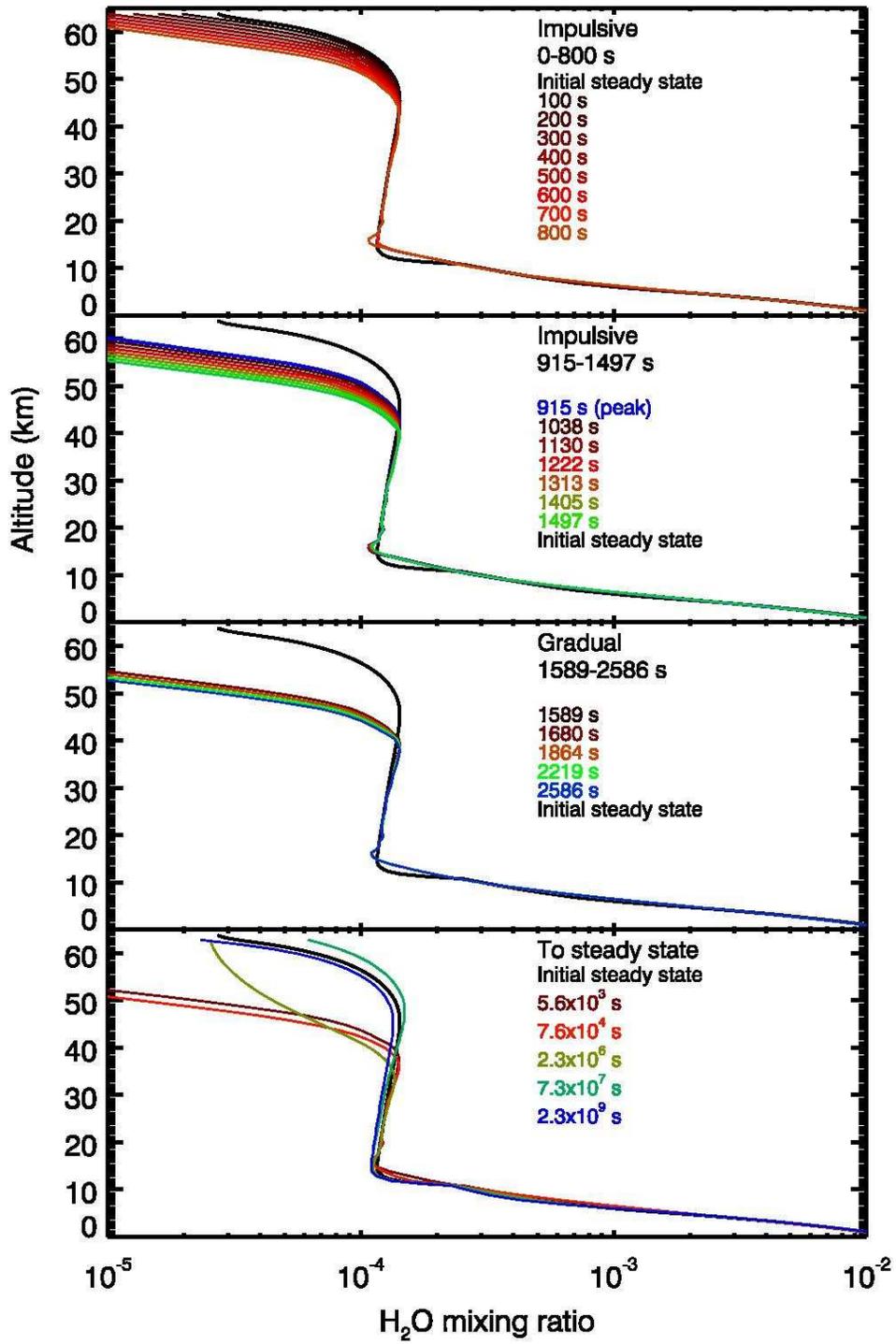

Figure 5. Water profile for an Earth-like planet around AD Leo, before, during, and after a big UV flare event.

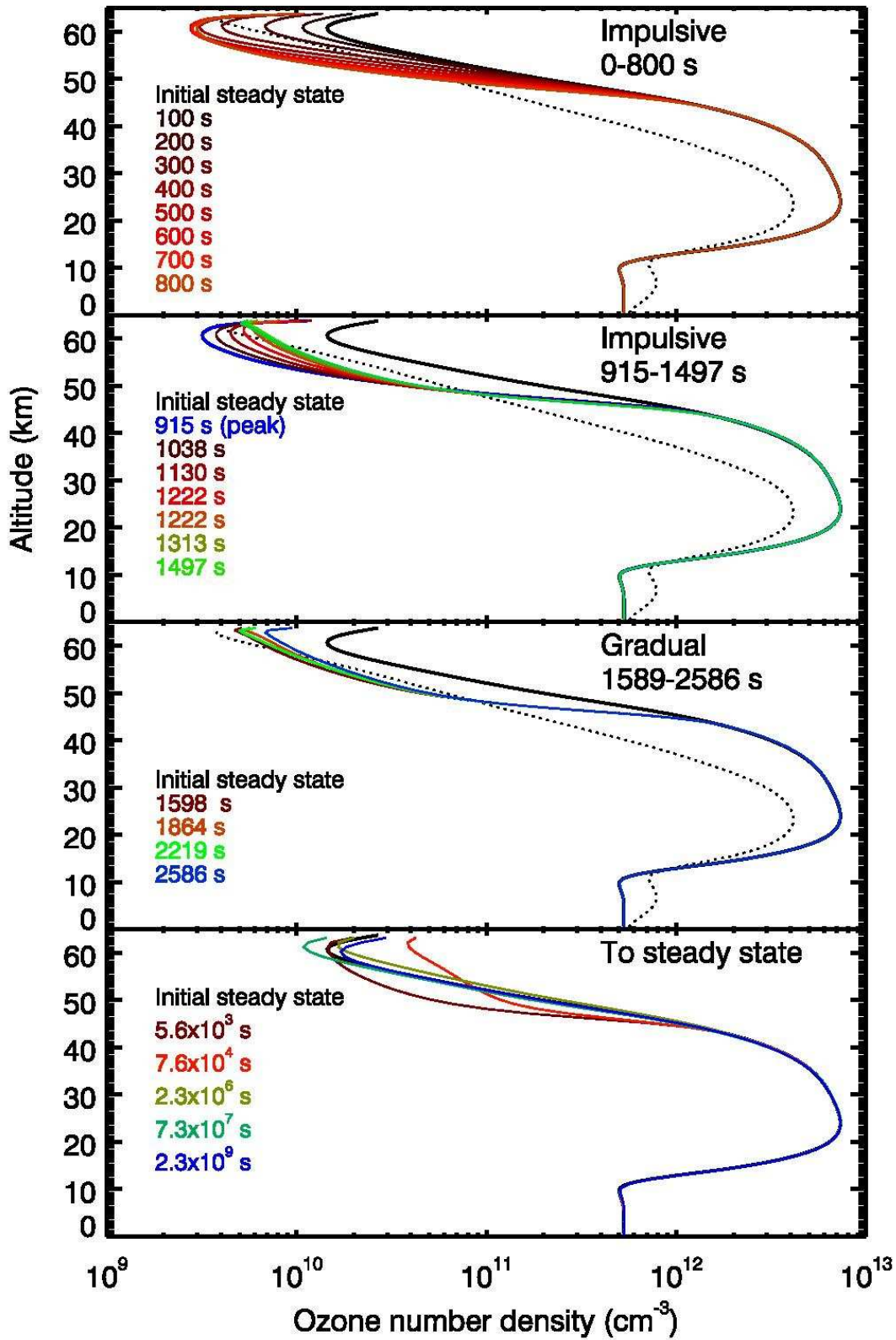

Figure 6. Ozone number density for an Earth-like planet around AD Leo, before, during, and after a big UV flare event. The ozone concentration calculated for Earth is presented for comparison (dotted line).

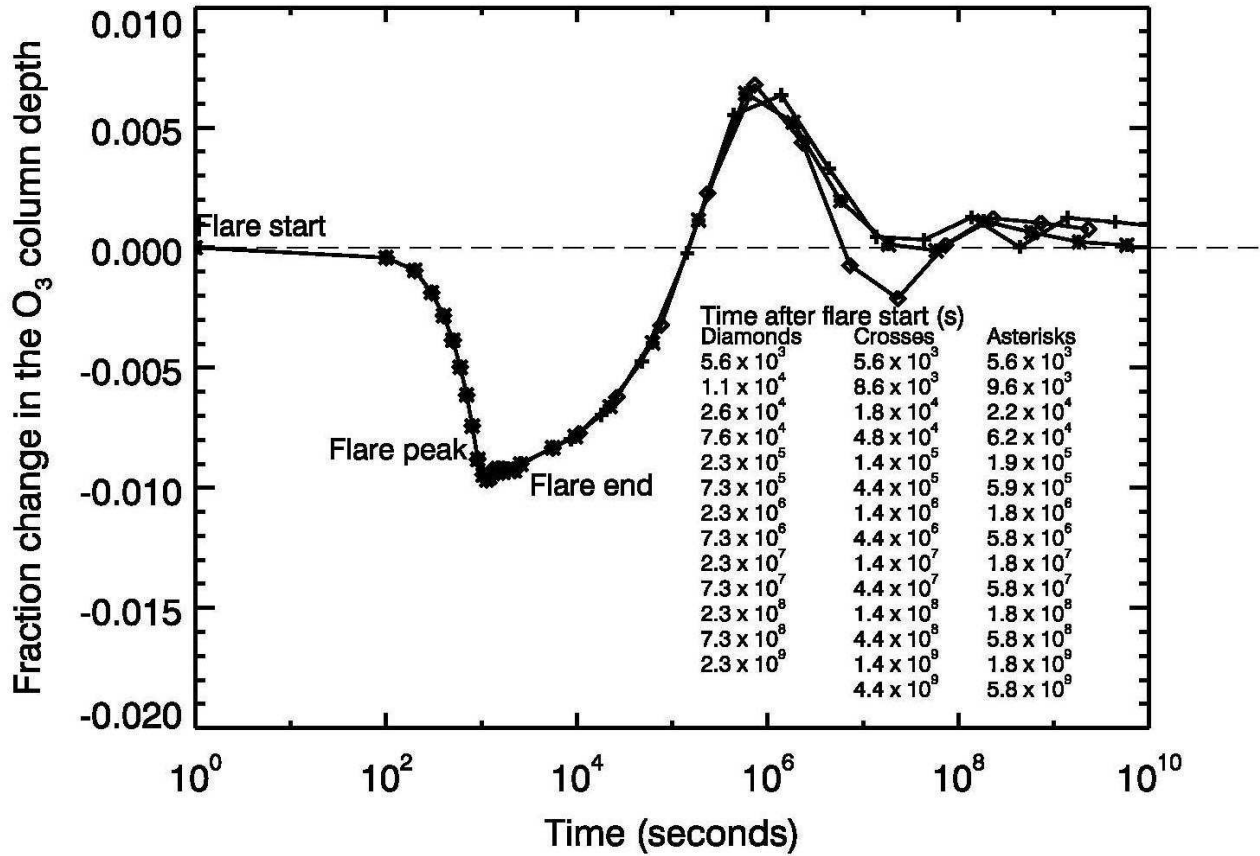

Figure 7. Time evolution of the ozone column depth compared to the initial steady state before, during, and after a big UV flare event. The lines show simulations made with different time steps after the flare ended. Times used for each run are listed in the figure.

*Atmospheric response to a proton event during a UV flare*
In this study we explored the effect on atmospheric chemistry and the planetary surface radiation environment for a flare that included both a UV component and a proton event. Figure 8 shows the evolution of the nitric oxide during the flare. At 915 s (flare peak) we introduced an increment to the NO abundance as explained in the *Input Proton Flux and Nitrogen Oxides production* section, to represent the effect of a large fluence of protons. After that, NO chemistry was allowed to evolve. When the effects of both UV radiation and proton flux are considered, the ozone density is depleted during the flare down to altitudes as low as 30km; after the flare the depletion becomes severe, reaching an altitude of 10 km from the ground (Fig. 9). The change in the ozone column depth is presented in Fig. 10. At the maximum depletion, the $O_3$ column depth was $1.1 \times 10^{18}$ cm$^{-2}$. This is 15 times lower than the initial $O_3$ column depth for the AD Leo planet and 7.5 times lower than the $O_3$ column depth calculated for present Earth by our model. When the peak of the proton event was delayed, the ozone depletion was about 10 percent lower than that calculated when both phenomena (UV flare and proton event) reach their maxima at the same time (Fig. 10). The difference between the two cases diminishes to 6 percent by $5.7 \times 10^6$ s and by $10^9$ seconds it is less than 1%. The lower ozone concentrations produced by the flare may still be detectable by missions like Terrestrial Planet Finder or Darwin, given that an ozone column depth as low as $3 \times 10^{17}$ cm$^{-2}$ produces a potentially detectable feature in the mid-IR planetary spectrum (Table 1, Fig. 13a in Segura et al. 2003).

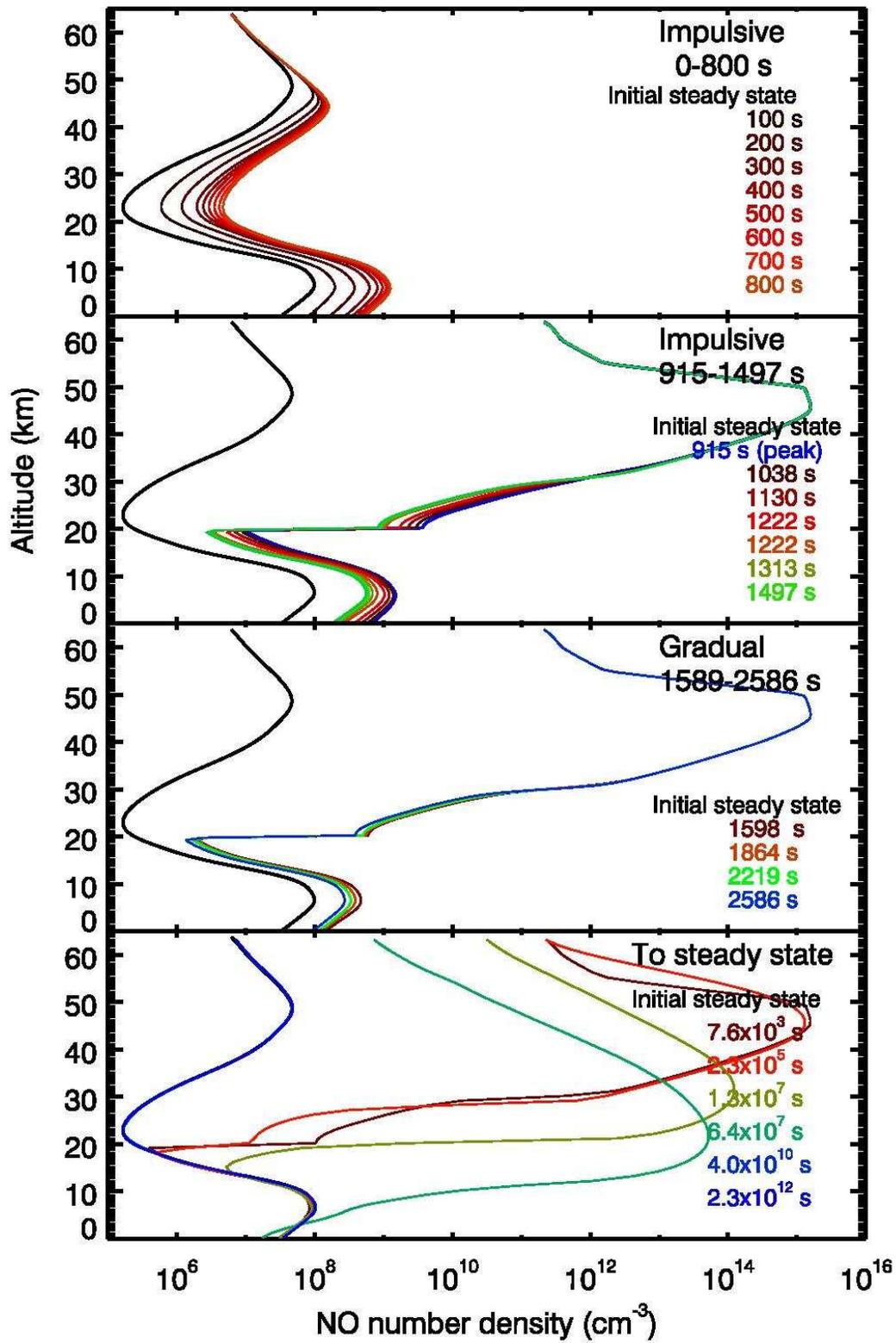

Figure 8. Time evolution before, during and after the flare for the atmospheric nitric oxide abundance profile. These results show the combined influence of the flare's incident UV radiation and a proton event at the peak of the flare.

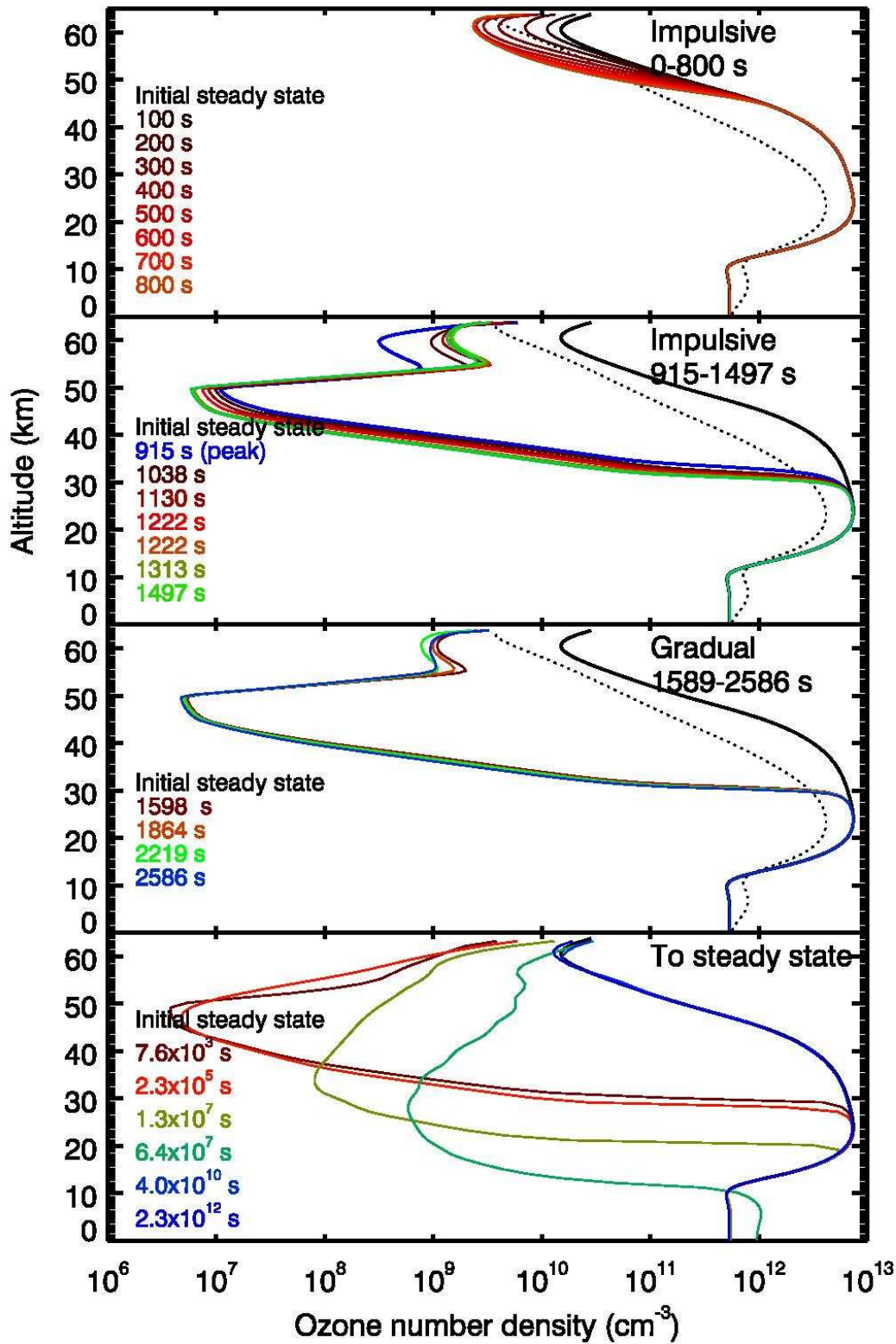

Figure 9. Time evolution before, during and after the flare for the atmospheric ozone abundance profile. These results show the combined influence of the flare's incident UV radiation and a proton event at the peak of the flare. The ozone concentration calculated for Earth is presented for comparison (dotted line).

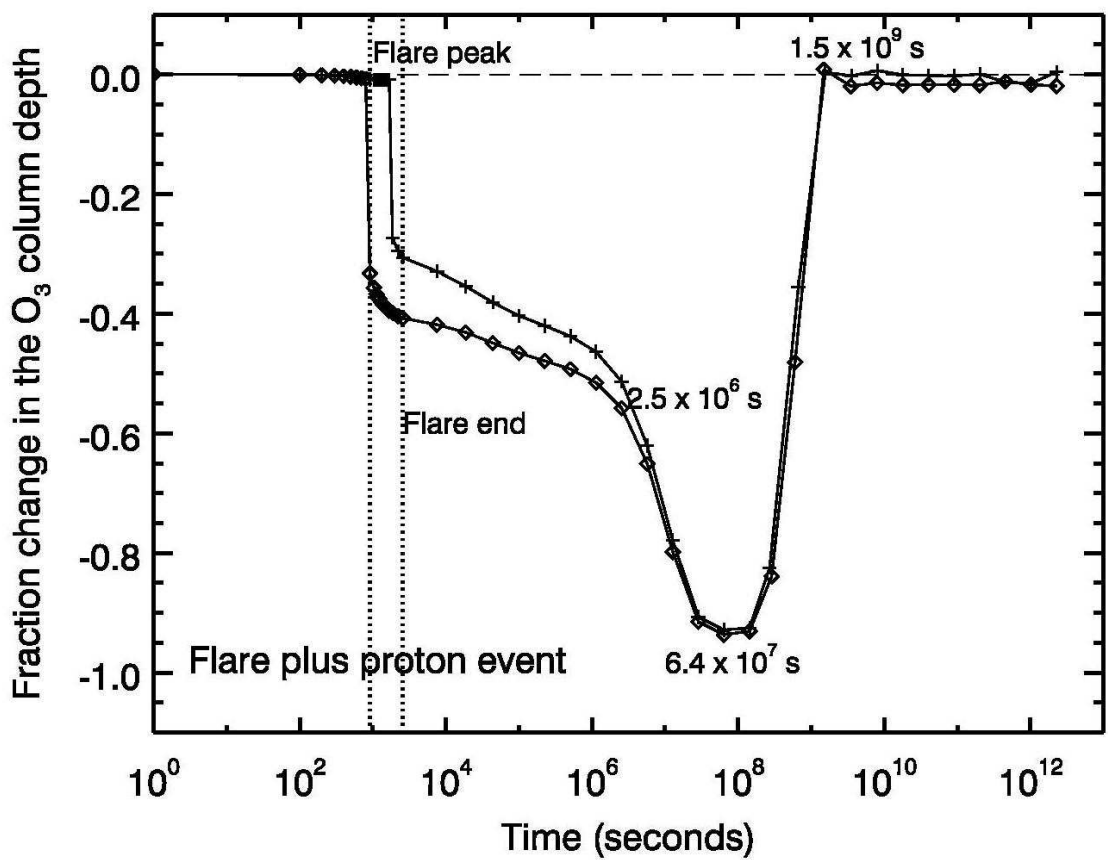

Figure 10. Time evolution of the ozone column depth compared to the initial steady state. These results show the combined influence of the flare's incident UV radiation and a proton event at the peak of the flare. Line with diamonds: O3 fraction change for a simultaneous UV and proton flux peak. Line with crosses: O3 fraction change for a proton event with a maximum delayed by 889 s with respect to the UV flare peak. Vertical dotted lines indicate the time for the peak of the UV flare and the end of the UV flare.

**Discussion**
As modeled here, an Earth-like planetary atmosphere is robust in response to a high-energy stellar flare if there are no associated charged particles. When only the UV flare radiation was considered, the modeled atmospheric temperature change was negligible at the surface, and did not exceed 6 K in the stratosphere during the flare, which is comparable to the measured midlatitude diurnal temperature variation on the Earth, for altitudes around 55 km (Huang et al. 2006). The tropospheric methane abundance remained unaffected, a behavior that was expected, given that the photochemical lifetime of methane for the simulated atmosphere is ~1000 years because of the low production of OH, the main sink of methane by the reactions described in the section *Planetary atmosphere simulation.* During the big AD Leo flare, the OH number density increases, reaching a maximum of $10^4$ cm$^{-3}$, but only for about 10 minutes and it returns to its number density values during quiescence right after the flare ends. The flare did produce a large drop in stratospheric water vapor and a change in the ozone column depth as the flare progressed. Water photolysis occurs from 1760 to 2050 Å, and is shielded by ozone. As $O_3$ becomes depleted at altitude above 40km, water at these altitudes is more readily photolyzed. Analysis of time-dependent results from the model indicates that direct photolysis of ozone in the upper stratosphere caused the initial drop in ozone number density during the impulsive phase of the flare. After the peak of the flare there are fewer incident photons to dissociate ozone, and recombination reactions dominate over photolysis, generating more ozone. The increase in ozone after the flare is caused by the large number of O atoms produced from $O_2$ dissociation at high altitudes during the flare. Once the flare ends, these atoms flow down into the upper stratosphere and recombine with $O_2$ to form $O_3$, resulting in a maximum in ozone density around $10^6$ s (~12 Earth days) following the flare (Fig. 7). This enhancement slowly re-equilibrates to the pre-flare value $10^7$ s after flare onset (over ~4 Earth months) but because of the very small magnitude of the enhancement, is unlikely to provide significant additional protection from subsequent flares during that time period.

The modeled UV flare is a particularly energetic, long duration flare for AD Leo, and can be considered representative of the highest energy flares likely to be encountered by a terrestrial planet orbiting an M dwarf. M dwarfs are variable over timescales from minutes to days, with shorter, less energetic flares being more common than strong, long-lasting flares. Typical flares on AD Leo range between those with total U band energies of $10^{30}$ ergs every few hours to $10^{32}$ erg events occurring every few days, with the mean flare energy being ~$4 \times 10^{30}$ ergs (Lacy, Moffett & Evans 1976), roughly 1000 times less energetic than the flare we used in our model .The frequency and energy distribution of flares on AD Leo are typical of very active M dwarfs, and are comparable to observations of ¡similarly active stars of its kind, such as YZ CMi (dM4.5e) and EV Lac (dM3.5e). As flare studies tend to focus on the most active stars (on which a given observing run is most likely to successfully observe a flare), little is known about the frequency or energetics of flares on intermediate- to low-activity M dwarfs. Activity at some level seems to be ubiquitous among M dwarfs— recent results from Sloan Digital Sky Survey repeat imaging find that flares occur even on relatively inactive cool stars (Kowalski et al. 2009). Our model therefore represents a limiting case, as we expect most flares will result in even lower UV flux at the planetary surface than calculated here.

During flare activity, the UV radiation incident upon a planet not only increases, but also changes its spectral slope. As Segura et al. (2005) show, the slope of the UV radiation emitted by the parent star is an important factor in a planet's atmospheric chemistry. As the absorption cross sections of the different atmospheric chemical species are wavelength dependent, the relative amount of near (2000 – 3000 Å) and far (1000 – 2000 Å) UV flux strongly affect the balance of chemical species in the atmosphere. The atmospheric chemistry is relevant for the kind and amount of chemical species that we may detect with instruments like Terrestrial Planet Finder (TPF) and Darwin (e.g. see Fig. 5, Segura et al. 2005). Flares of different energies emit UV with different slopes: in the far UV flat and positive slopes seem to dominate (e.g. Fig. 7, Bromage et al. 1986) while the near UV emitted by flares has only been measured a few times (Hawley and Pettersen 1991, Jevremović et al. 1998) making general behavior difficult to infer. For the present experiment, the far UV flare spectra had to be interpolated for time steps between the actual measurements, resulting in flat-slope far UV spectra, similar to observations of other M dwarfs like AU Mic, Gl 825 and Gl 867 A (Butler et al. 1981). The resulting constant slope during the flare (Fig. 2) may not be characteristic for all flares, however—van den Oort et al. (1996) observed a flare on YZ CMi where the far UV slope changed as the flare evolved. We will explore the effect of this slope change on planetary atmospheric chemistry in a future study.

*Surface UV and biological damage*

For the purpose of estimating the biological damage, the near UV is divided in three regions: UV-C (<2800 Å), UV-B (2800–3150 Å) and UV-A (3150-4000 Å). When the proton event is not included, the UV flux that reaches the surface at the flare peak is 50 times larger than the initial surface flux during stellar quiescence, but this increase takes place at wavelengths longer than 3150 Å (Fig. 11). UV-A radiation is roughly 100 times less effective at damaging DNA relative to shorter wavelength radiation (Segura et al. 2003, 2005). Comparing the UV-A surface flux at the peak of the flare with the UV-A flux received on Earth's surface we found that the AD Leo planet surface receives only 8% more energy than Earth does in this region of the UV spectrum. This increment lasts for about 300 s.

Energetic particles affect the atmospheric chemistry by promoting the formation of nitrogen oxides that destroy ozone; as a result the planetary surface is less protected from the UV radiation that comes from the parent star (Fig. 12). Table 2 presents the values of UV irradiation for the top of the atmosphere and planetary surface for selected times before, during and after the flare with the effect of the proton event included. As can be seen on Table 2, the surface UV flux at the AD Leo planet is larger than the one received by present Earth at the peak of the flare in the UV-A and UV-B wavelength ranges, and at $6.4\times10^6$ s after the flare in the UV-C range.

We calculated the UV dose rates by convolving the surface UV fluxes with an action spectrum for DNA damage (Van Baalen and O'Donnel, 1972) as was done by Segura et al. (2005). For most of the time this UV dose is smaller than that for Earth. At the flare peak and at the time of the maximum ozone depletion the UV dose rate for DNA damage is 14 and 4 percent larger, respectively, than that on Earth.

It is important to remember that flares are localized phenomena on the stellar surface. The radiation and particles emitted during a flare will reach the planet only if they are directed towards the planet. The calculations done here for a proton event are an upper limit, not only because of the extremely high fluence, but because not all flares have associated particle ejections and, when a flare does emit energetic ionizing particles, they may not impact the planet. This is particularly true of planets that have magnetic fields (see below).

*Limitations of 1-D models*
Earth has been subjected to solar flares throughout its geologic history. We have no record of any solar flare as energetic as those emitted by the most active M dwarfs, but still the measurements and models of Earth's atmosphere response during these events provide clues as to how a planetary atmosphere recovers after a stellar flare. Studies of Earth's atmosphere's response to solar flares have generally focused on the effect of particles and ionizing radiation (X-rays and EUV). These studies show that atmospheric winds play a key role in allowing atmospheric composition to recover to average values by redistributing chemical species in the atmosphere. Also, the Earth's magnetic field only allows the entrance of ionized particles through the poles, so the most drastic effects on the atmospheric chemistry are seen at high latitudes. The global average of the ozone depletion caused by these events is low. For example, 2-D and 3-D models that simulated the effect of the "Carrington Event" solar flare show that most of the ozone depletion occurs at high latitudes, and the maximum ozone depletion occurred two months after the flare when $NO_x$ species were transported from the upper atmosphere to the lower stratosphere (Thomas et al. 2007). Globally averaged, this depletion reached 5%, similar to the currently observed global depletion caused by human activities. While the average depletion was not high, the local values of ozone depletion for high latitudes reached 14% two months after the flare. The atmosphere came back to normal values about 4 years after the flare.

Because our model is 1-D, our simulations correspond to a planet with no magnetic field, which receives the ionizing particles in the hemisphere that faces its parent star at the moment of the event. The predicted recovery times inferred from our simulations are around $10^9$ s (~50 years). In order to calculate detailed ozone depletion caused by the particles emitted by the flare, 2-D or 3-D models may be required, depending on whether the planet has a magnetic field.

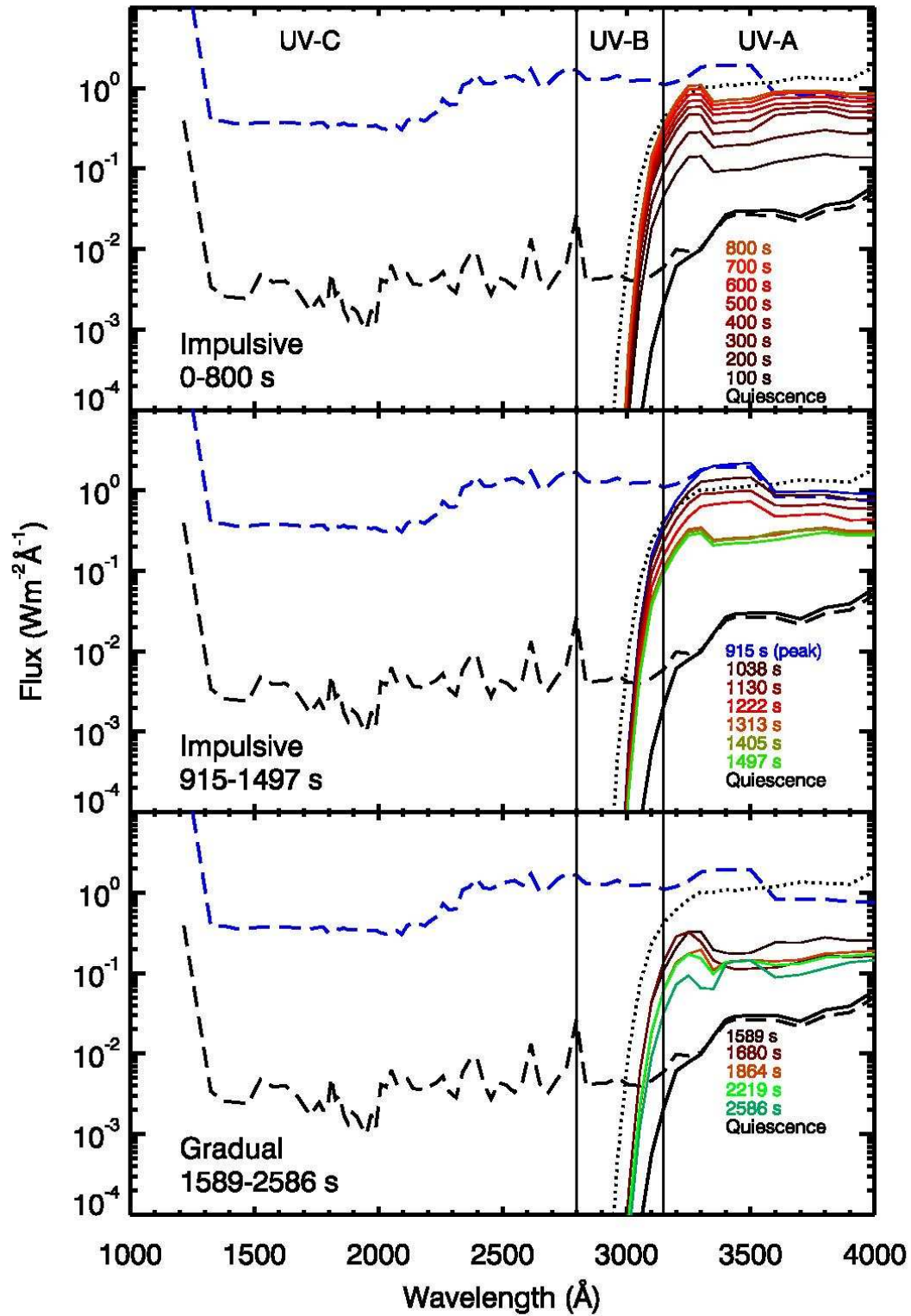

Figure 11. Surface UV flux before, during and after the UV flare (solid lines). The top of the atmosphere UV flux at quiescence (dashed black line) and at the peak of the flare (dashed blue line) are drawn for comparison. The dotted line is the surface UV radiation calculated for Earth. Vertical solid lines show the boundaries between the regions known as UV-A (3150-4000 Å), UV-B (2800-3150 Å), and UV-C (< 2800 Å).

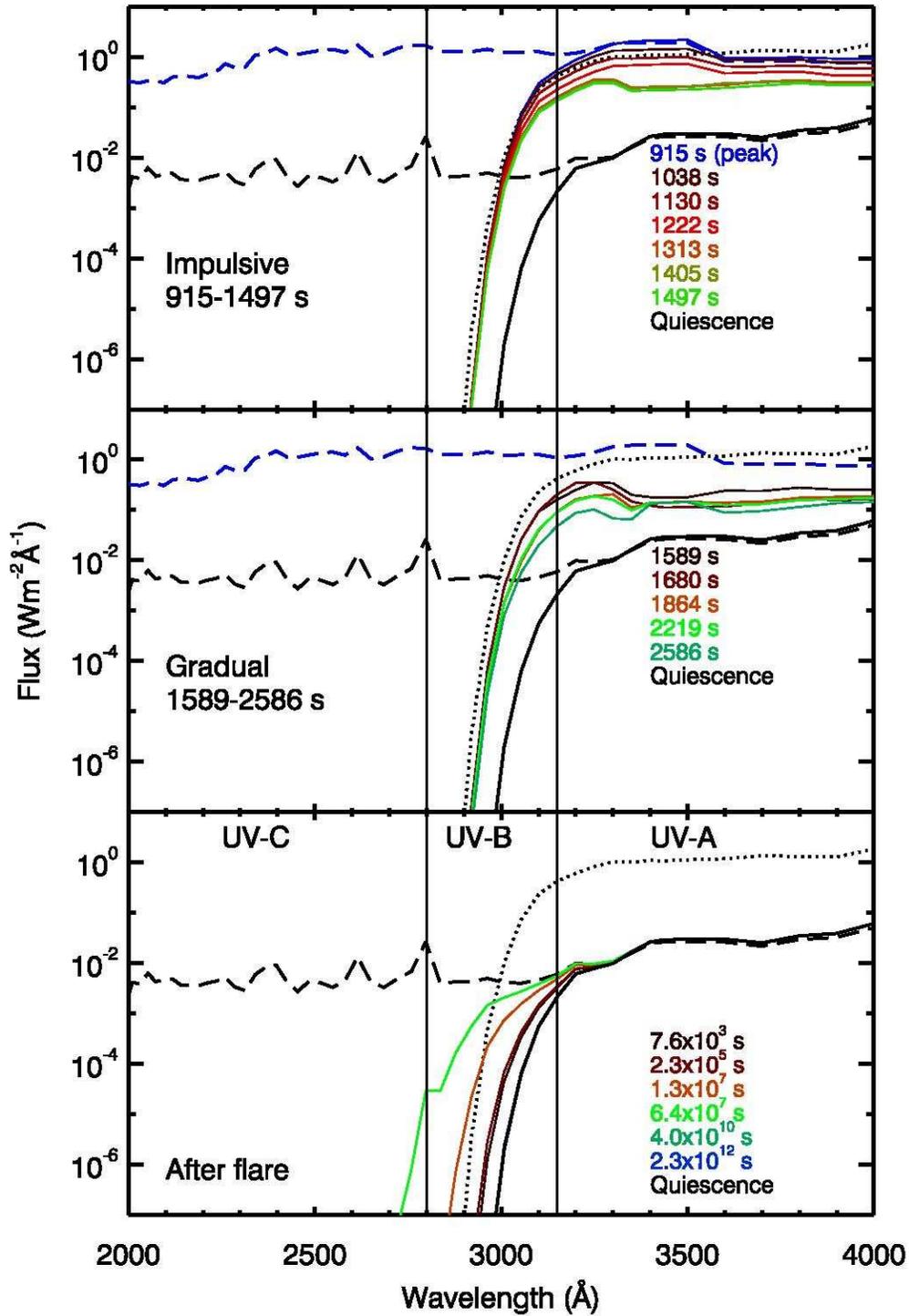

Figure 12. Surface UV flux before, during an UV flare plus proton event and after the flare (solid lines). The top of the atmosphere UV flux at quiescence (dashed black line) and at the peak of the flare (dashed blue line) are drawn for comparison. The dotted line is the surface UV radiation calculated for Earth. Vertical solid lines show the boundaries between the regions known as UV-A (3150-4000 Å), UV-B (2800-3150 Å), and UV-C (< 2800 Å). The first 800 seconds of the flare are omitted here because fluxes are identical to those shown in Fig. 11.

Table 2. Ultraviolet integrated flux in W/m$^2$ for selected times before, during and after the UV flare with a proton event included. Earth values are shown for comparison. (TOA= Top of the atmosphere)

| | UV-A (3150-4000 Å) | | UV-B (2800-3150 Å) | | UV-C (<2800Å) | |
|---|---|---|---|---|---|---|
| | TOA | Surface | TOA | Surface | TOA | Surface |
| Earth | 102.36 | 118.45 | 17.23 | 2.55 | 6.73 | $2.13 \times 10^{14}$ |
| AD Leo planet | | | | | | |
| Quiescence (t=0 s) | 2.60 | 2.97 | 0.20 | 0.01 | 2.76 | $2.13 \times 10^{-14}$ |
| Flare start (t=100 s) | 10.89 | 11.59 | 5.34 | 0.21 | 43.10 | $1.93 \times 10^{-14}$ |
| Flare peak (t= 915 s) | 112.17 | 120.77 | 45.43 | 3.15 | 368.76 | $1.93 \times 10^{-14}$ |
| After flare (t=7.6× 10$^3$ s) | 2.60 | 3.00 | 0.20 | 0.02 | 2.76 | $1.93 \times 10^{-14}$ |
| After flare (t=1.3× 10$^7$ s) | 2.60 | 3.02 | 0.20 | 0.04 | 2.76 | $2.52 \times 10^{-10}$ |
| After flare (t=6.4× 10$^7$ s) | 2.60 | 3.03 | 0.20 | 0.06 | 2.76 | $5.85 \times 10^{-5}$ |
| After flare (t=1.4× 10$^8$ s) | 2.60 | 3.03 | 0.20 | 0.06 | 2.76 | $3.38 \times 10^{-5}$ |
| After flare (t=6.0× 10$^8$ s) | 2.60 | 3.00 | 0.20 | 0.02 | 2.76 | $1.93 \times 10^{-14}$ |

**Conclusions**

For an oxygen-rich, Earth-like planet in the habitable zone of an active M dwarf, stellar flares do not necessarily present a problem for habitability. Much of the potentially life-damaging UV radiation goes into photolyzing ozone in the stratosphere, preventing it from reaching the planetary surface. Ozone variations cause temperature fluctuations in the upper atmosphere, but these fluctuations are small, and the climate at the surface is unaffected. Ionizing particles emitted during a flare may be more dangerous depending on how much of the particle flux strikes the planet. The additive effects of repeated flares over the duration of the planet's lifetime are not well understood-- as M dwarfs can be active on timescales of days to weeks, the atmosphere may not return to equilibrium before another flare occurs.

The changes in the ozone column depth and the lack of variation of methane concentrations indicate that one energetic flare should not impede the detection of biogenic compounds by instruments like TPF or Darwin.

**Acknowledgements**

A.S. and J.F.K. acknowledge the support of the NASA Terrestrial Planet Finder -Foundation Science Program via solicitation NRA NNH04ZSS001N. This work was also performed as part of the NASA Astrobiology Institute's Virtual Planetary Laboratory Lead Team, supported by the NASA Astrobiology Institute under Cooperative agreement No. CAN-00-OSS-01, and solicitation NNH05ZDA001C. A. S. acknowledges the support from the CONACYT projects No. 51715 and 79744, and project PAPIIT IN119209-3. L.M.W. and S.L.H. acknowledge the support of HST program 10525, provided by NASA through a grant from the Space Telescope Science Institute, which is operated by the Association of Universities for Research in Astronomy, Inc., under NASA contract NAS 5-26555. S.L.H. acknowledges support from NSF grant AST 0807205.